\documentclass[a4paper,11pt]{article}
\usepackage[utf8]{inputenc}
\usepackage{lmodern}
\usepackage[T1]{fontenc}
\usepackage{amsmath,amsthm,amsfonts,amssymb,bm}
\usepackage{dsfont}			
\usepackage{enumitem}
\usepackage{booktabs}
\usepackage{xcolor}
\usepackage{authblk}
\usepackage{appendix}
\usepackage[authoryear]{natbib}
\setcitestyle{square}
\usepackage[right=2.5cm,left=2.5cm,top=2cm,bottom=3cm]{geometry}
\usepackage[font={small,it}]{caption}	
\usepackage[colorlinks,linkcolor=blue,citecolor=blue,urlcolor=blue]{hyperref}
\usepackage{subfig}
\usepackage{mathtools}

\usepackage{multirow}
\usepackage{makecell}
\usepackage{afterpage}
\usepackage{footnote}
\usepackage{float}
\usepackage{graphicx}
\usepackage{caption}

\title{ \textsc{Classification of cow diet based on milk Mid Infrared Spectra: a data analysis competition \\ at the ``International Workshop of \\ Spectroscopy and Chemometrics 2022''} }
\author[1,2]{Maria Frizzarin}
\author[3]{Giulio Visentin\footnote{Corresponding author: address. Email: giulio.visentin@unibo.it}}
\author[4]{Alessandro Ferragina}
\author[5]{Elena Hayes}
\author[6]{Antonio Bevilacqua}
\author[6]{Bhaskar Dhariyal}
\author[7]{Katarina Domijan}
\author[4]{Hussain Khan}
\author[6]{Georgiana Ifrim}
\author[6]{Thach Le Nguyen}
\author[2,8]{Joe Meagher}
\author[9]{Laura Menchetti}
\author[6]{Ashish Singh}
\author[2,8]{Suzy Whoriskey}
\author[10]{Robert Williamson}
\author[11]{Martina Zappaterra}
\author[12]{Alessandro Casa}

\affil[1]{Teagasc, Animal \& Grassland Research and Innovation Centre, Moorepark, Ireland}
\affil[2]{School of Mathematics and Statistics, University College Dublin, Ireland}
\affil[3]{Department of Veterinary Medical Sciences, University of Bologna, Italy}
\affil[4]{Teagasc Food Research Centre, Ashtown, Ireland}
\affil[5]{Teagasc, Food Research Centre, Moorepark, Ireland}
\affil[6]{School of Computer Science, University College Dublin, Ireland}
\affil[7]{Department of Mathematics and Statistics, National University of Ireland, Maynooth, Ireland}
\affil[8]{Insight Centre for Data Analytics, University College Dublin, Ireland}
\affil[9]{School of Biosciences and Veterinary Medicine, University of Camerino, Italy}
\affil[10]{School of Electronics, Electrical Engineering and Computer Science, Queen's University Belfast, UK}
\affil[11]{Department of Agricultural and Food Sciences, University of Bologna, Italy}
\affil[12]{Faculty of Economics and Management, Free University of Bozen-Bolzano, Italy}

\date{}                     

\begin{document}

\maketitle

\begin{abstract}
In April 2022, the Vistamilk SFI Research Centre organized the second edition of the ``International Workshop on Spectroscopy and Chemometrics – Applications in Food and Agriculture''. Within this event, a data challenge was organized among participants of the workshop. Such data competition aimed at developing a prediction model to discriminate dairy cows’ diet based on milk spectral information collected in the mid-infrared region. In fact, the development of an accurate and reliable discriminant model for dairy cows’ diet can provide important authentication tools for dairy processors to guarantee product origin for dairy food manufacturers from grass-fed animals. Different statistical and machine learning modelling approaches have been employed during the workshop, with different pre-processing steps involved and different degree of complexity. The present paper aims to describe the statistical methods adopted by participants to develop such classification model. 
\end{abstract}


\section{Introduction}
The use of mid-infrared spectroscopy (MIRS) has become a relevant topic in agri-food sciences, due to its capacity to routinely quantify a wide range of important characteristics rapidly and cost-effective. In particular, MIRS is nowadays commonly employed to monitor and quantify milk quality parameters, such as concentrations of fat, protein, casein, and lactose. These parameters are used for milk quality-based payment schemes, genetic and genomic selection, and as farmers’ support tool. Spectral information generated from MIRS analysis have also proven to be effective in predicting fine milk quality parameters, including protein fractions, free amino acids \citep{bonfatti2011effectiveness,mcdermott2016prediction}, individual and groups of fatty acids \citep{soyeurt2006estimating, fleming2017prediction}, milk processing traits \citep{ferragina2013use,visentin2015prediction}, animal-related characteristics \citep{mcparland2014mid,shetty2017predicting,ho2019classifying}, and can be used as a tool for the verification of the authenticity of agricultural foods \citep{cozzolino2012recent}. A more extended list of applications of MIRS in the dairy science framework can be retrieved from the reviews by \citet{de2014invited} and \citet{tiplady2020evolving}.

The two-day event \emph{``International Workshop on Spectroscopy and Chemometrics''} was organized by Vistamilk SFI Research Centre in April 2022, following its first edition held in 2021 \citep{Frizzarin2021MidIS}. The workshop focused on describing the main challenges and applications of near and mid-infrared spectroscopy in food, animal, and agricultural sciences with internationally recognised researchers. Moreover, participants, on a voluntary basis, were provided with a large dataset containing individual cow milk spectra with the sole information on animal's diet for a chemometric data competition. Such data presented many challenges from a methodological and statistical point of view, due to the high dimensionality of the spectral matrices, and strong collinearity between adjacent spectral wavelengths. The chemometric challenge, therefore, encouraged the engagement of participants with different background and skills and required the application of different statistical and machine learning strategies. 

The purpose of the data challenge was to develop a model to predict the diet fed to dairy cows by exploiting mid-infrared spectral information. Participants, or groups of participants, were required to apply their developed model to a test set containing only individual milk spectra and to submit their prediction of animals’ diet. Although the participation to the chemometric challenge was extremely high among participants, only the best six contributions, in terms of accuracy of prediction and methodological innovativeness, were selected to present their results both at the workshop and in the present manuscript.

\section{Data description and challenge}\label{sec:sec2}
A dataset consisting of 4,364 individual milk spectra from 120 cows was collected between May and August in 2015, 2016 and 2017 \citep{o2016effect}. The samples were from Holstein Friesian cows with different parity from Irish Dairy Research Herd in Teagasc Moorepark, Fermoy, Co. Cork. Three dietary groups were evaluated with 54 cows being assigned to each dietary group each year. The three diet treatments were grass (GRS) which consisted of perennial ryegrass only, clover (CLV) which consisted of perennial ryegrass with 20\% annual clover sward, and total mixed ration (TMR) where cows were fed grass silage, maize silage and concentrates while being maintained indoors for the full season. Milk samples were collected in the morning (AM) and evening (PM) milking session; subsequently AM+PM samples were pooled and analysed weekly using Pro-FOSS FT6000 (FOSS). A total of 1060 transmittance data points in the region from 925 cm$^{-1}$ to 5,000 cm$^{-1}$ were collected. 

The dataset was divided into training (3275 spectra) and test (1089 spectra) data; for the latter only spectral information was provided, while diet information, to be used as a classification variable, was available for the training set. The training data included 1094 spectra for GRS, 1120 spectra from CLV and 1061 spectra for TMR.  There were no missing values in the training or test set. The specific information about the wavenumbers had not been shared with the participants.

The three dietary groups were carefully selected based on their characteristics. As described by \citet{frizzarin2021application}, pasture-based diets are easily discriminated from TMR diets, while discriminating between GRS and CLV diets is much more difficult due to the similarities in the sward composition resulting in similar milk composition. However, with the increased pressure to reduce fertilizer use, and the introduction of multi-species swards, the development of a robust discriminant model for classifying milk spectra based on diet is of paramount importance.

After the analysis, the participants submitted their predicted values for the test dataset and a short explanation of the methodology used. The best methods were selected based on the novelty of the contribution and on the accuracy of the predictions for the test dataset. The accuracy was calculated as the proportion of the correctly classified samples divided by the total number of samples in the test dataset. 

\section{Modelling approaches and results}\label{sec:sec3}

\subsection{Participant 1}\label{sec:tabular}

The data were analyzed following different modelling strategies, focusing both on methods that considered the ordering of the wavelengths and on methods that do not. All the analyses have been mainly conducted using \texttt{Python} libraries \texttt{pandas, sklearn, sktime} and \texttt{matplotbib} \citep[see][and references therein]{pedregosa2011scikit}: the code is available at \url{https://github.com/mlgig/vistamilk\_diet\_challenge}. 

As a first step, some descriptive statistics were computed, and the outliers have been removed, following both the recommendations given prior to the competition and a visual inspection of the data. In the subsequent step, the labeled dataset was split according to a 3-fold cross-validation (3CV) strategy.  Therefore, the best model was selected based on cross-validation accuracy, and then trained on the full training set and used to perform prediction on the provided unlabeled test set.

In order to predict the diet, the following classification strategies were considered: 
\begin{itemize}
    \item {\bf Tabular models:} each sample is considered as a vector of unordered features. In particular, Ridge Classifier and Linear Discriminant Analysis (LDA) were tested. In the following, these methods were coupled both with feature selection strategies and with random polynomial feature transformations. The latter approach, by generating new polynomial variables from the original ones, aimed to check if non-linear interactions improved the classification accuracy. In particular, a new approach is presented which aimed to diversify polynomial features while keeping low computational requirements. 
    \item {\bf Deep Neural Network Models:} a family of approaches based on deep neural networks, both fully connected and convolutional, were tested. This strategy implicitly generates complex features interactions, as captured by the network architecture.
\end{itemize}

Note that previously obtained results \citep{Frizzarin2021MidIS} suggest that tabular methods work quite well with spectroscopy data. Moreover, following the suggestions in \citet{frizzarin2021application}, feature selection strategies were coupled with the information about the presence of water regions in the spectra. In addition, state-of-the-art time series classification algorithms, such as ROCKET \citep{dempster2019rocket}, MiniROCKET \citep{minirocket}, MrSQM \citep{mrsqm1, mrsqm2} and FreshPrince \citep{freshprince}, were tested. Lastly, \emph{ensemble methods} were applied, aiming to mix together time series and tabular models, to combine their predictions and strengths. Nonetheless, these approaches have been outperformed by the ones mentioned above, therefore the corresponding results are not shown in the next sections. 

\subsubsection{Tabular models, feature selection and transformation}\label{sec:tach}
In Table \ref{tab:tabular_results}, results for the best tabular methods are presented. Both the ridge classifier, appropriately tuned, and LDA performed quite well, while being extremely fast to train. Nonetheless, the selection of some specific wavelengths seemed to improve the accuracy further. In fact, both the removal of the noisy water regions and the data-driven feature selection (performed using the \texttt{SelectFromModel} routine in Python), provides better results. 

\begin{figure}[t]
    \centering
    \includegraphics[width = 15cm, height = 8cm]{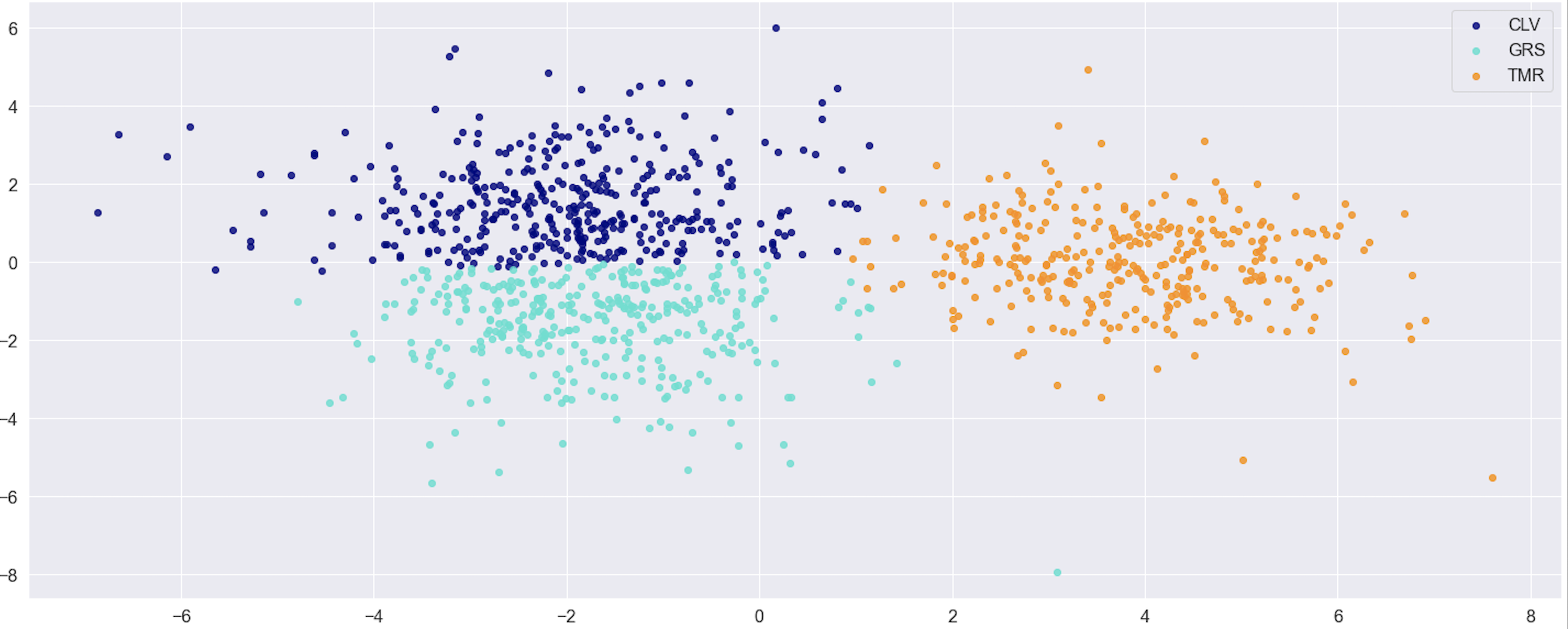}
    \caption{LDA visualisation for the model 
Feature Selection + Polynomial Features + 
LDA, applied to the unlabeled test data to predict class labels.}
    \label{fig:lda-poly2}
\end{figure}

Nevertheless, all these approaches hover around 80\% accuracy, therefore, in order to improve it, the data were augmented considering polynomial features of degree two (using \texttt{sklearn} method \texttt{PolynomialFeatures(degree = 2)}). This led to an increase of the accuracy to 84.4\%. The LDA component visualisation for the model with Feature Selection and Polynomial Features, applied on the unlabeled test dataset, is shown in Figure \ref{fig:lda-poly2} and a good discrimination between the three classes is clearly visible.  

\begin{table}[t]
    \centering
    \small
    \caption{Accuracy results, evaluated on the 3-fold cross-validation, for the tabular methods considered, coupled with feature selection strategies.}
\begin{tabular}{lc}
\hline
\textbf{Method} &  \textbf{Accuracy} \\\hline
Ridge Classifier & 0.760 \\
LDA & 0.747  \\
Feature Selection + Ridge Classifier & 0.777   \\                                   Feature Selection + LDA  & 0.778    \\
No water + Ridge Classifier & 0.777 \\
No water + LDA & 0.783 \\
Feature Selection + Polynomial Features + LDA & 0.844 \\
No water + Feature Selection + Polynomial Features + LDA & 0.844 \\
\hline
\end{tabular}
    \label{tab:tabular_results}
\end{table}

The improvements obtained when considering polynomial features, come at a price in terms of the computational requirements. In fact, starting from the 1060 original wavelengths, the addition of second-degree polynomial features resulted in a total number of variables which made the model estimation task unfeasible. To address this issue, in this work a new \emph{Random Polynomial Features} (\texttt{RPolyTransformer} in the following) approach was introduced. The key idea was to implement random sampling in the non-linear feature space. This lead to relevant advantages as the total number of features can be controlled and it can consider both higher-degree ($>2$) polynomial features and complex mathematical functions (e.g., cosine, exp). 

This strategy firstly generated $K$ random arithmetic expressions (see Table \ref{tab:rpoly_example} for some examples), which are then used to compute $K$ non-linear features. From the new and the original features, $K^*$ variables are selected using \texttt{SelectKBest} from \texttt{sklearn}. The hyperparameters $K$ and $K^*$ were optimized via cross-validation in the final model (see the final row of Table \ref{tab:rpoly_results}).

\begin{table}[tb]
    \centering
    \small
    \caption{Examples of RPolyTransformer features used. Here $x_j$ denote the $j$-th wavelength.}
\begin{tabular}{l}
\hline
 \\
$(x_{32} * x_{19}) + x_{103} - x_{2} $  \\
$(x_{102} * (x_{78}) + x_{26}) $  \\
$(x_{1} - x_{150}) + x_{64} * x_{4} * x_{5} $  \\
  \\
\hline
\end{tabular}
    \label{tab:rpoly_example}
\end{table}

\begin{table}[]
    \centering
    \small
    \caption{Results for different combinations with \texttt{RPolyTransformer}. \texttt{SelectFromModel} and \texttt{SelectKBest} are feature selection modules to remove noise from data (the former) and select  the most discriminative non-linear features (the latter).}
\begin{tabular}{p{13cm}c}
\hline
\textbf{Method} & \textbf{Accuracy} \\
\hline
Region: FULL &  \\
& \\
RPolyTransformer + Ridge Classifier  & 0.717 \\ 
RPolyTransformer + LDA & 0.619 \\
SelectFromModel + RPolyTransformer + SelectKBest + LDA & \textbf{0.848}  \\
\hline
Region: {[}925:1585, 1720:2989{]} & \\
& \\
RPolyTransformer  + Ridge Classifier & 0.805 \\ 
RPolyTransformer  + LDA  & \textbf{0.847} \\
SelectFromModel + RPolyTransformer + SelectKBest + LDA   & 0.843  \\
\hline
Region: {[}925:1585, 1720:2989, 3738:3807{]} & \\
& \\
RPolyTransformer  + Ridge Classifier & 0.811 \\ 
RPolyTransformer  + LDA    & 0.833 \\
SelectFromModel + RPolyTransformer + SelectKBest + LDA   & 0.835   \\
\hline
\textbf{Optimized model}  &  \\
Region: {[}925:1585, 1720:2989{]} & \\
RPolyTransformer($K = 17000$)  + SelectKBest($K^* = 7000$) + LDA & \textbf{0.864} \\ 
\hline
\end{tabular}
    \label{tab:rpoly_results}
\end{table}

In Table \ref{tab:rpoly_results} the results obtained with this method, again combined with different classifiers and feature selection approaches and tested with the full data and the data after water region removal, are presented. 
At first, when combining \texttt{RpolyTransformer} with a classifier, a significant drop in the accuracy was observed, if compared with simple tabular models. Ridge was more accurate than LDA but it was still far behind the previous results. However, by carefully filtering the features either automatically with \texttt{SelectFromModel} or manually by removing the water regions, the results improved noticeably. In these experiments, LDA outperforms Ridge consistently. Compared to the \texttt{PolynomialFeatures} method, the one proposed here is faster (a few seconds versus a few minutes) and just as accurate. However, the initial results without noise reduction (i.e., feature selection) suggest that this strategy is more sensitive to noise in the data. 

\subsubsection{Deep Learning Models}\label{sec:deepLearningGeorg}
When considering deep learning models, the task of exploding the feature space and learning feature interactions is completely deferred to the network, without requiring any feature engineering steps. In turn, deep neural networks require a careful design process, to avoid overfitting and to identify the best model architecture and input modality.

\begin{figure}[t]
	\centering
	\begin{minipage}{.5\textwidth}
		\textsf{\hspace{2cm}Full sequence (33x33)}\par\medskip
		\centering
		\includegraphics[width=1\linewidth]{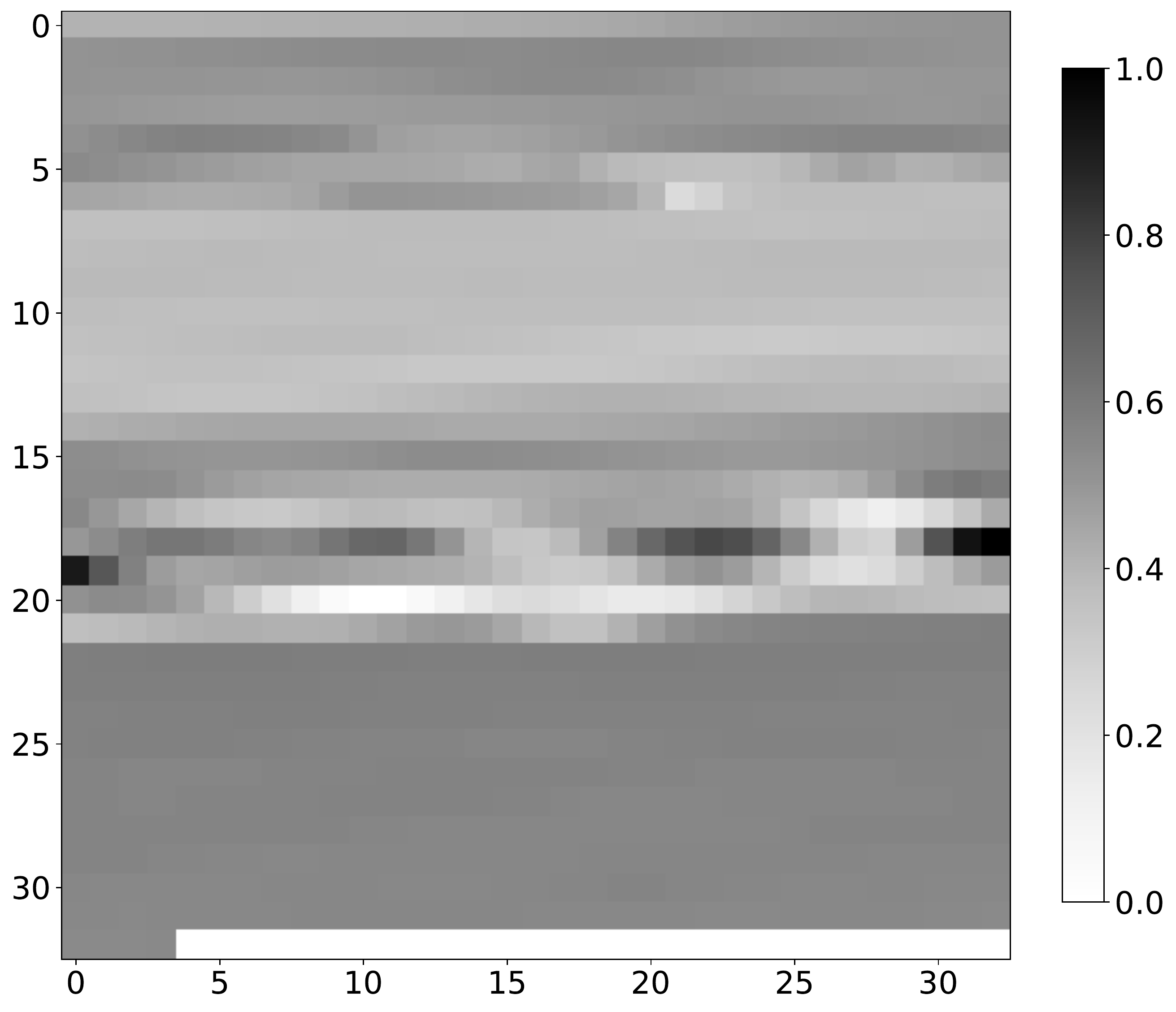}
	\end{minipage}%
	\begin{minipage}{.5\textwidth}
 \textsf{\hspace{1.5cm} No water regions (23x23)}\par\medskip
		\centering
		\includegraphics[width=\linewidth]{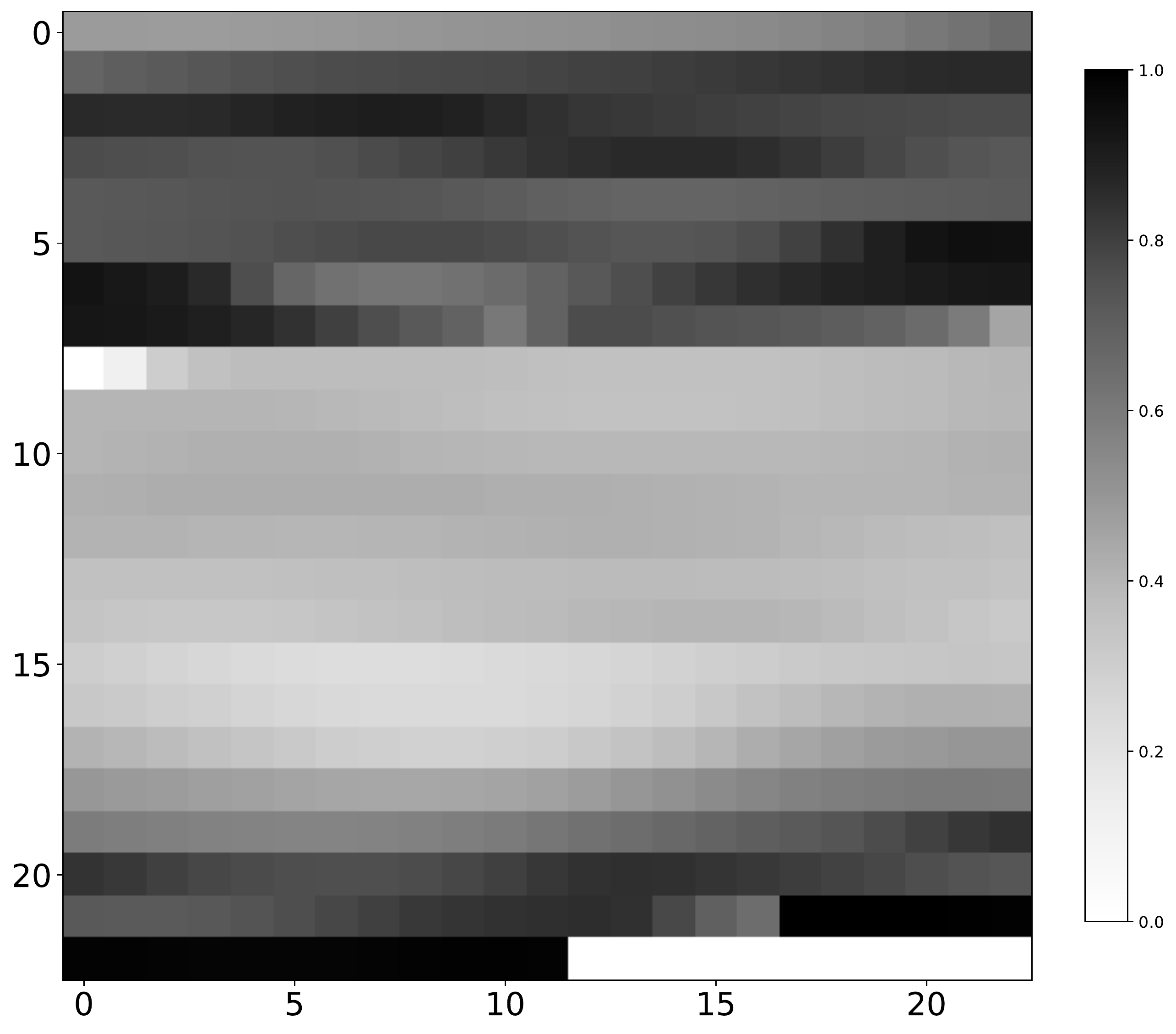}
	\end{minipage}
	\caption{Spectroscopy sequences arranged as image structures. In both examples, the padding values are visible at the bottom of the resulting images. Values are normalised in the 0-1 range for convenience.}
	\label{imagelike}
\end{figure}

The designed model architectures considered here can be grouped into two main categories, namely, Fully Connected Networks (FCNs) and Convolutional Neural Networks (CNNs). FCNs do not require any manipulation or adaptation of the input data, as each single wavelength is treated as an independent feature and fed to an input unit. In contrast, CNNs require the data to be bi-dimensional, image-like matrices, as they are commonly used to address image classification problems. For this family of networks, the input waves need then to be vertically stacked as 2D arrays and therefore, in order to fit the closest squared dimension, padded with trailing zeros. An example of how the spectroscopy sequences can be presented to the CNNs is provided in Figure \ref{imagelike}. Additionally, a third group of models is tested for this challenge, namely, CNNs based on dilated kernels (further denoted as CNN\_DILATED). Whilst regular CNNs extract features through compact squared filters, or local receptive fields, the CNN\_DILATED network utilizes filters that are spatially dilated by a fixed factor \citep{yu2015multi}.  Dilated kernels are commonly used in semantic image segmentation.

All the models in this group were trained on both the full training dataset and on the water reduced one. When the CNN models were trained, the full data were shaped into images of shape 33x33 with a padding of 29 values, while the reduced data were shaped into images of shape 23x23 with a padding of 11 values. As already mentioned, all padding values were zeros, and they were appended to the original sequences.

The full list of the implemented architectures is presented in Table \ref{deeparch} in Appendix \ref{sec:appendixA}. The experiments were conducted on the previously described 3-fold cross-validation splits; note that, for each split, 20\% of the training data was held back for validation purposes, to identify network hyperparameters such as number of training epochs, initial learning rate, or regularisation rates. Models were trained for a total of 50,000 epochs, with an early stopping policy used to monitor the validation loss to detect overfitting and save time during the training phase. The final model used to classify the provided unknown data was selected as the overall best performing architecture, and trained over the full training data for a number of epochs set as the average of the epochs reached during the 3CV training.

All models were implemented using \texttt{TensorFlow} \citep{tensorflow2015-whitepaper}, and trained on a workstation featuring a single GPU, model Nvidia Titan XP. Results are presented in Table \ref{table:cvresults}, which contains the training performances obtained over the 3-folds CV experimental campaign. For all the tested architectures, excluding the water regions from the input waves resulted in a performance increase of roughly 12-13\%. The FCN model working on data after water-region removal, achieved the highest accuracy across the 3 splits, with an average of 84.7\%. Similar unreported results were obtained also considering a single split validation strategy, which furthermore demonstrated that convolutional models tend to overfit the input data quite fast. 

\begin{table}[]
\centering
\caption{Training results on the 3CV splits. }
\begin{tabular}{llccccc}
\hline
\textbf{Model}                & \textbf{Data} & \textbf{Split 1} & \textbf{Split 2} & \textbf{Split 3} & \textbf{Average} \\ \hline
\multirow{2}{*}{FCN}          & FULL          & 0.670             & 0.677           & 0.675           & 0.674           \\
                              & NO WATER      & \textbf{0.854}           & \textbf{0.851}            & \textbf{0.837}           & \textbf{0.847}           \\ \hline
\multirow{2}{*}{CNN}          & FULL          & 0.686  & 0.684  & 0.670  & 0.680  \\
                              & NO WATER      & 0.806           & 0.836           & 0.832           & 0.824           \\ \hline
\multirow{2}{*}{CNN\_DILATED} & FULL          & 0.678           & 0.684           & 0.652           & 0.671           \\
                              & NO WATER      & 0.824           & 0.812           & 0.807           & 0.814           \\ \hline
\end{tabular}
\label{table:cvresults}
\end{table}

\subsection{Participant 2}\label{sec:hussein}
All the processing steps and the algorithm implementation was completed using \texttt{MATLAB} \citep{MATLAB}. After having imported the dataset in tabular form, the outliers were identified as those observations with more than three scaled median absolute deviations from the median of the dataset. Classification was performed using a set of algorithms such as Support Vector Machine (SVM), K-Nearest Neighbors (KNN) and Linear Discriminant Analysis (LDA).  Hyperparameters tuning and evaluation of the classification accuracy were performed via 5-fold cross-validation. 

\begin{figure}
    \centering
    \includegraphics[scale = 0.8]{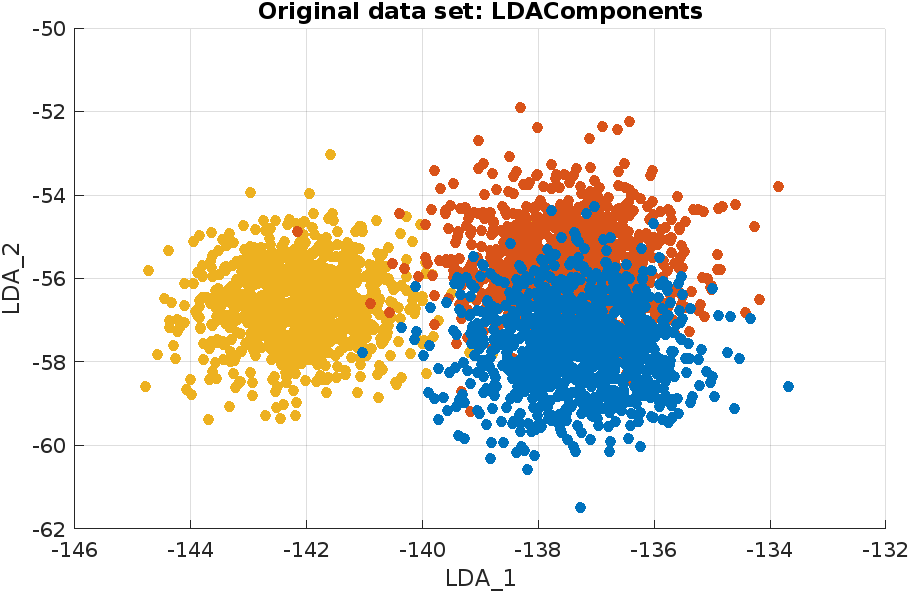}
    \caption{LDA components extracted from the developed model.}
    \label{fig:hussein}
\end{figure}

The best results were obtained using LDA, which was able to distinguish outdoor grass-feed cow’s milk from TMR with an accuracy of 95\% while differentiating grass and clover with an accuracy of 68\%. Figure \ref{fig:hussein} allows to visualize class boundaries by plotting the spectra projections in the latent space spanned by the two discriminant functions. From the figure, a clear boundary can be observed between the indoor and outdoor feed classes, while there is a significant overlap between the GRS and CLV classes. Therefore, the extracted components were then considered as an input to a linear SVM model to improve classification between outdoor feed classes. The combination of two classifier (LDA + SVM), resulting in a two-step approach,  significantly improved the overall classification accuracy (87.1\%) as well as classification accuracy between classes, as shown in Table \ref{tab:hussein}. 

\begin{table}[t]
    \centering
    \caption{Confusion matrix obtained by combining LDA and SVM.}
\begin{tabular}{cc| ccc}
& & & Predicted class & \\
 & & CLV & GRS & TMR  \\
\hline
& CLV & 83.5\% & 17.4\% & 0.7\% \\
True class & GRS & 15.8\% & 81.6\% & 0.8\% \\
& TMR & 0.7\% & 1.1\% & 98.5\%  \\

\end{tabular}
    \label{tab:hussein}
\end{table}

\subsection{Participant 3}\label{sec:group_bologna}
The present work was developed independently by three group members, following a common preliminary analysis of spectral data. Results of the prediction on the test set provided for the chemometric challenge were then compared to assess the agreement between the three different statistical approaches employed.

\subsubsection{Preliminary edits on spectral data}
These edits were conducted on raw spectral data in both the training and test sets using \texttt{Python}. Spectra expressed in transmittance were converted into absorbance by taking the $\log_{10}$ of the reciprocal of the transmittance. Subsequently, spectral wavelengths associated to water absorption, as well as non-informative regions, were deleted. This led to a reduced version of the dataset, that has been used for the subsequent analyses, with 511 remaining wavelengths in the regions between 2,994 and 1,682 cm$^{-1}$ and between 1,578 and 926 cm$^{-1}$. A graphical representation of this procedure is reported in the supplementary material (Figure \ref{fig:figUniBoSupplementary}).

\subsubsection{First approach}\label{sec:memb1_unibo}
To explore the multivariate structure of the dataset, Principal Component Analysis (PCA) was exploited on the training dataset, using \texttt{prcomp} function in \texttt{stats} package and the \texttt{factoextra} package \citep{factoextra} in the \texttt{R} environment \citet{R}. The analysis revealed that most of the data variability was explained by the first two Principal Components (PCs), accounting together for the $88\%$ of the total variance (see the scree plot on the left top panel in Figure \ref{fig:figUniBo}). 

Afterwards, possible outliers were detected using the algorithm proposed by \citet{filzmoser2008outlier} and implemented in the \texttt{mvoutlier} package \citep{mvoutlier}; only the observations being both location and scatter outliers were removed from the training dataset. As a results, a total of 63 observations were removed from the training dataset.


After outliers removal, linear discriminant analysis was considered using \texttt{lda} function in the \texttt{MASS} package \citep{MASSpackage}. To test its accuracy, as a first step the discriminant functions were applied to the training dataset, with the aim of comparing the estimated classification with the actual one. Therefore, LDA was first applied to maximize the differences between TMR and the CLV+GRS (in the following named PAST group). The LDA returned one Linear Discriminant (LD) function, which was then applied to the training dataset to attribute the TMR diet to observations. Afterwards, LDA was applied again by maintaining in the training set only the observations belonging to the PAST group. The obtained LD function was then applied to the whole training dataset to discriminate between CLV and GRS diets previously categorized as PAST. The vector with the predicted classes was then compared with the vector of actual group classification in the training dataset, thus computing the training accuracy. This appproach resulted in an overall model training accuracy equal to 83.3\% (see Table \ref{tab:unibo}); the scatter plot of the first versus second linear dimension scores is depicted in the right top panel in Figure \ref{fig:figUniBo}. Lastly, the LD functions obtained on the training dataset allowed for the classification of the unknown observations in the test dataset, with the results reported in Table \ref{tab:unibo}.

\begin{table}[t]
    \centering
    \small
    \caption{Summary of the results of the three different approaches.}
\begin{tabular}{p{3.6cm}| p{3.3cm}p{3.7cm}p{3.7cm}}
\hline
& Member 1 & Member 2 & Member 3 \\
\hline
Brief description & Two steps DA in \texttt{R} & Canonical DA with stepwise method in \texttt{SAS} & DA with stepwise methods in \texttt{SPSS} \\
& & & \\
Number of samples (training set) & 3180 & 3116 & 3153 \\
Number of wavelengths retained & 511 & 88 & 16 \\
Accuracy (training set) & 83.30\% & 81.32\% & 71\% \\
\hline
\end{tabular}
\vspace{0.2cm}\\
\textbf{Predicted diet for the samples in the test dataset (n cases)} \\
\begin{tabular}{p{3.6cm}| p{3.3cm}p{3.7cm}p{3.7cm}}
\hline
TMR & 344 & 326 & 365 \\
CLV & 367 & 342 & 326 \\
GRS & 366 & 353 & 386 \\
\hline
\end{tabular}
\vspace{0.2cm}\\
\textbf{Agreement between the approaches applied to the test dataset} \\
\begin{tabular}{p{3.6cm}| p{3.3cm}p{3.7cm}p{3.7cm}}
\hline
Member 1 &  &  &  \\
Member 2 & 84.21\% &  &  \\
Member 3 & 72.90\% & 70.84\% &  \\
\hline
\end{tabular}
    \label{tab:unibo}
\end{table}

\begin{figure}[ht]\label{fig:figUniBo} 
  \begin{minipage}[b]{0.5\linewidth}
    \centering
    \includegraphics[height = 5cm, width = 7cm]{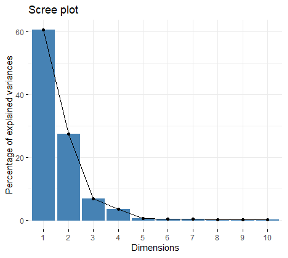}
    \vspace{4ex}
  \end{minipage}
  \begin{minipage}[b]{0.5\linewidth}
    \centering
    \includegraphics[height = 5cm, width = 7cm]{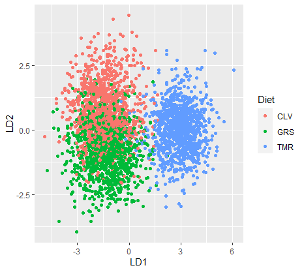}
    \vspace{4ex}
  \end{minipage} 
  \begin{minipage}[b]{0.5\linewidth}
    \centering
    \includegraphics[height = 5cm, width = 7cm]{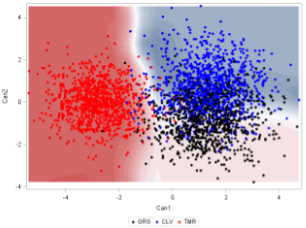}
    \vspace{4ex}
  \end{minipage}
  \begin{minipage}[b]{0.5\linewidth}
    \centering
    \includegraphics[height = 5cm, width = 7cm]{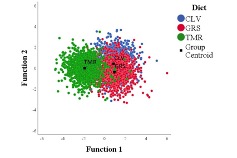}
    \vspace{4ex}
  \end{minipage} 
  \caption{Explained variance by the first 10 principal components (top left), scatter plot of discriminant models developed by member 1 (right top), member 2 (bottom left) and member 3 (bottom right).}
  \label{fig:figUniBo} 
\end{figure}

\subsubsection{Second approach}\label{sec:memb2_unibo}
Principal component analysis (\texttt{PROC PRINCOMP}, SAS Institute Inc., ver. 9.4) was undertaken on the training set, as in Section \ref{sec:memb1_unibo}. Coherently, outlier removal was then performed by calculating the Mahalanobis distance (MD) as the uncorrected sum of squares of the first four centred and scaled PC scores, explaining up to the 98.21\% of the total spectral variance. Outliers were defined as samples whose MD was greater than the 97.5th percentile of a $\chi^2$ distribution with 4 degrees of freedom \citep{brereton2015mahalanobis}. Following this approach, a total of 127 samples were discarded from the training set.

The discriminant model was developed following a multiple-step approach. Firstly, a stepwise discriminant analysis was carried out in order to identify the most significant wavelengths associated with the three different diets using the \texttt{PROC STEPDISC}. A total of 88 wavelengths were retained and used for the subsequent canonical discriminant analysis, which was developed through the \texttt{PROC DISCRIM}. The proportion of samples correctly classified was 73.38\% (CLV), 73.70\% (GRS), and 97.62\% (TMR), with an overall model accuracy of 81.32\%. The scatter plot of the first versus second canonical variables scores is in the bottom left panel of Figure \ref{fig:figUniBo}. The wavenumbers with the greatest (in absolute value) canonical discriminant function coefficients were between 1,154 and 1,162 cm$^{-1}$, 2,843 cm$^{-1}$, 2,874 cm$^{-1}$, and 2,882 cm$^{-1}$, thus providing some potentially relevant information to be explored to assess which milk chemical features are more influenced by the dietary regimen. The discriminant model was then applied to the test set to obtain the prediction of cows’ diet on unknown milk spectra.

\subsubsection{Third approach}\label{sec:memb3_unibo}
Standard assumptions required for multivariate analyses were verified before proceeding to the main analysis. Two diagnostic measures were used to identify the outliers for the predictors and the dependent variables; in the former case Mahalanobis Distance (MD) was used to spot multivariate outliers while, in the latter one, studentized residuals were considered. Samples whose MD was greather than the 97.5th percentile of the MD distribution and studentized residuals greater than 2.5 were removed. During this process, a total of 90 outliers have been identified and excluded. Potential multicollinearity was then verified by Tolerance and Variance Inflation Factors. Moreover, the ratio between the number of cases and predictors was checked as an indicator of the adequacy of the sample size; a ratio of 20 observations for each predictor variable, with the smallest group size exceeding the number of independent variables, is suggested \citep{meloun2011statistical,pituch2015applied}. 

LDA was then chosen as the main discriminative approach. The stepwise method, using Wilks’ lambda $\Lambda$ as criterion, was adopted to reduce multicollinearity and increase the case/predictors ratio, improving the adequacy of the sample size. Box’s test and log determinants were considered to verify the equality of covariance matrices. The canonical correlation and the proportion of between-group variance that is due to each variate were used as measures of effect size \citep{pituch2015applied}, while the performance of the LDA was evaluated by classification-related statistics and leave-one-out CV \citep{hahs2016applied}. The \texttt{Scoring Wizard} command was finally used to apply the discriminant functions (DF) to the test dataset, and the predicted probability was calculated to assess its performance. Analyses were performed with SPSS software \citep{SPSS}.

Standardized canonical DF coefficients of the variables selected by DA and measures of effect size are shown in Table \ref{tab:uniBoAppendix} in the Supplementary Material. More than 90\% of the total difference between the groups was attributable to the first DF, with the Wilks’ $\Lambda$ (0.330) indicating that it has a significant discriminating capacity (p-value < 0.001). Wavenumber 2,851 cm$^{-1}$ and 2,890 cm$^{-1}$ mostly contributed to the discrimination of cows’ diet. The second DF only explained 6\% of the total variance, being nonetheless still significant (Wilks’ $\Lambda$= 0.902; p-value < 0.001). Centroids (Table \ref{tab:uniBoAppendix2}) and the plot of DF scores (bottom right panel in Figure \ref{fig:figUniBo}) indicated that the first DF appropriately discriminate the TMR group from the others (i.e., CLV and GRS). On the other hand, group separation on the second DF was poor; in particular, CLV and GRS clusters were not clearly distinguished. The cross-validation procedure indicated an overall model accuracy of 71\% (see Table \ref{tab:unibo}), with different sensitivity between groups: over 90\% for TMR samples, and below 65\% for CLV and GRS samples. The application of DFs to predict the diet of cows in the test data set showed a similar trend, with an expected sensitivity of 64\%, 63\%, and 87\% for CLV, GRS, and TMR diets, respectively (Table \ref{tab:uniBoAppendix3}).

\subsection{Participant 4}\label{sec:robertwilliamson}
A conventional machine learning pipeline was used, composed of feature (i.e., wavelength) selection and classification, with no outliers being removed from the original dataset. Dimensionality reduction techniques such as Principal Component Analysis (PCA) and Independent Component Analysis (ICA), as well as Extended Multiplicative Scatter Correction (EMSC) and a data augmentation approach were tested to improve the classification results \citep{bjerrum2017data}. EMSC represents a preprocessing technique which removes multiplicative effects potentially caused by physical phenomena such as light scattering, which is commonly seen in reflectance spectroscopy, thus allowing for easier modelization of chemical effects. On the other hand, the data augmentation scheme increases the data set ten fold by adding random variations in offset, multiplication, and slope, nine times to each sample. 

Subsequently a range of different classifiers, which have successfully been adopted before on infrared spectroscopy data, were used. In particular, the considered models were K-nearest Neighbour \citep[K-NN;][]{balabin2011biodiesel}, Random Forest \citep[RF;][]{chen2021efficient}, Support Vector Classification \citep[SVC;][]{ji2013rapid}, Multilayer-perceptron \citep[MLP;][]{balabin2008gasoline}, Linear Discriminant Analysis \citep[LDA;][]{khuwijitjaru2020near}, Decision Tree Classification \citep{geronimo2019computer}, Nu-Support Vector Classification \citep[NuSVC;][]{terouzi2013classification}, AdaBoost Classification \citep{wu2017prediction}, Gradient Boosting Classification \citep{munera2021discrimination}, Gaussian Naive Bayes \citep{bhati2020iot} and Quadratic Discriminant Analysis \citep[QDA;][]{oravec2019forensic}.
Other investigated predictive methods belonged to the group of deep Learning (DL) techniques, and in particular one-dimensional (1D) Convolutional Neural Network (CNN). 1D CNN makes use of six one dimensional convolutional layers, and a number of max pooling, batch normalization and dropout layers. Each 1D CNN layer is followed by a max-pooling and batch normalization layer. One-dimensional CNN was apply only on raw spectra in order to retain the sequence of the data, not required for PCA and ICA. 

Prior to the analyses, the dataset was split in a training set (80\% of the data), to train the different models, and a validation set (remaining 20\% of the data), used to optimise the hyperparameters and to identify the best method to be used for final testing.

\begin{figure}[ht]
    \centering
    \includegraphics[scale = 0.75]{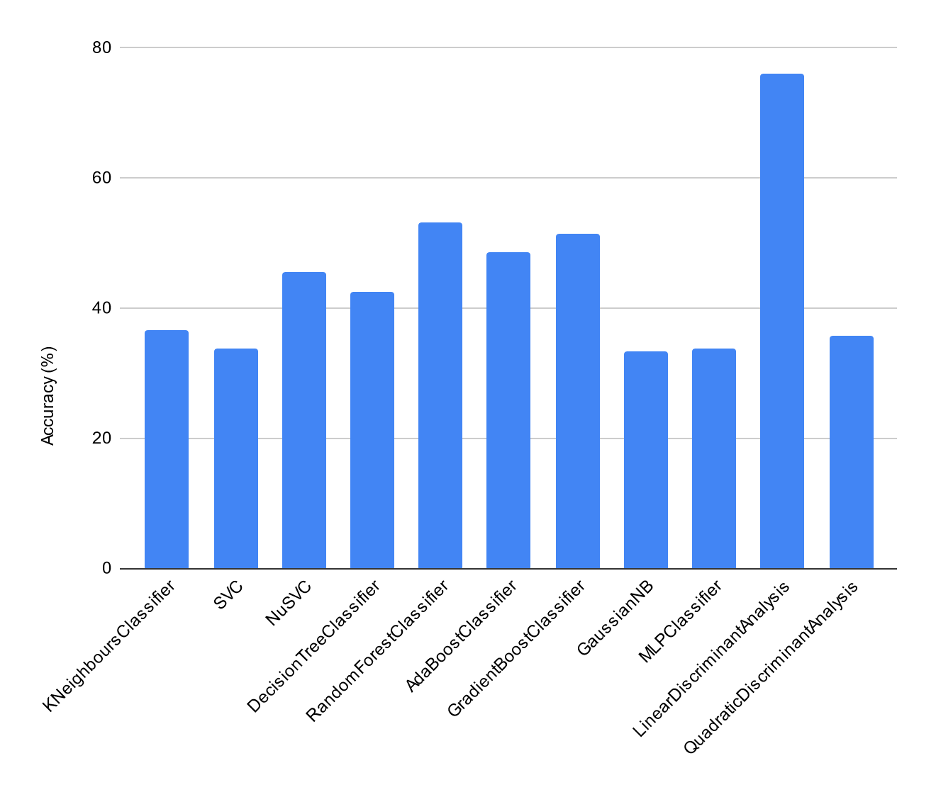}
    \caption{Results of classifiers on a 80/20 test-train split.}
    \label{fig:figRobert1}
\end{figure}

An initial experiment was performed on all classifiers without the use of data augmentation or feature selection. This was carried out to explore which classification method was performing better with the raw spectral data. Figure \ref{fig:figRobert1} shows the results obtained from the initial step with the 80/20 train/validation for different classifiers. All results gathered were averages taken from three training and validation predictions for each model. LDA gave the best results with an accuracy of 76\%, whereas the MLP and SVC produce some of the worst performances with accuracies around 33\%.

\begin{figure}[ht]
    \centering
    \includegraphics[scale = 0.8]{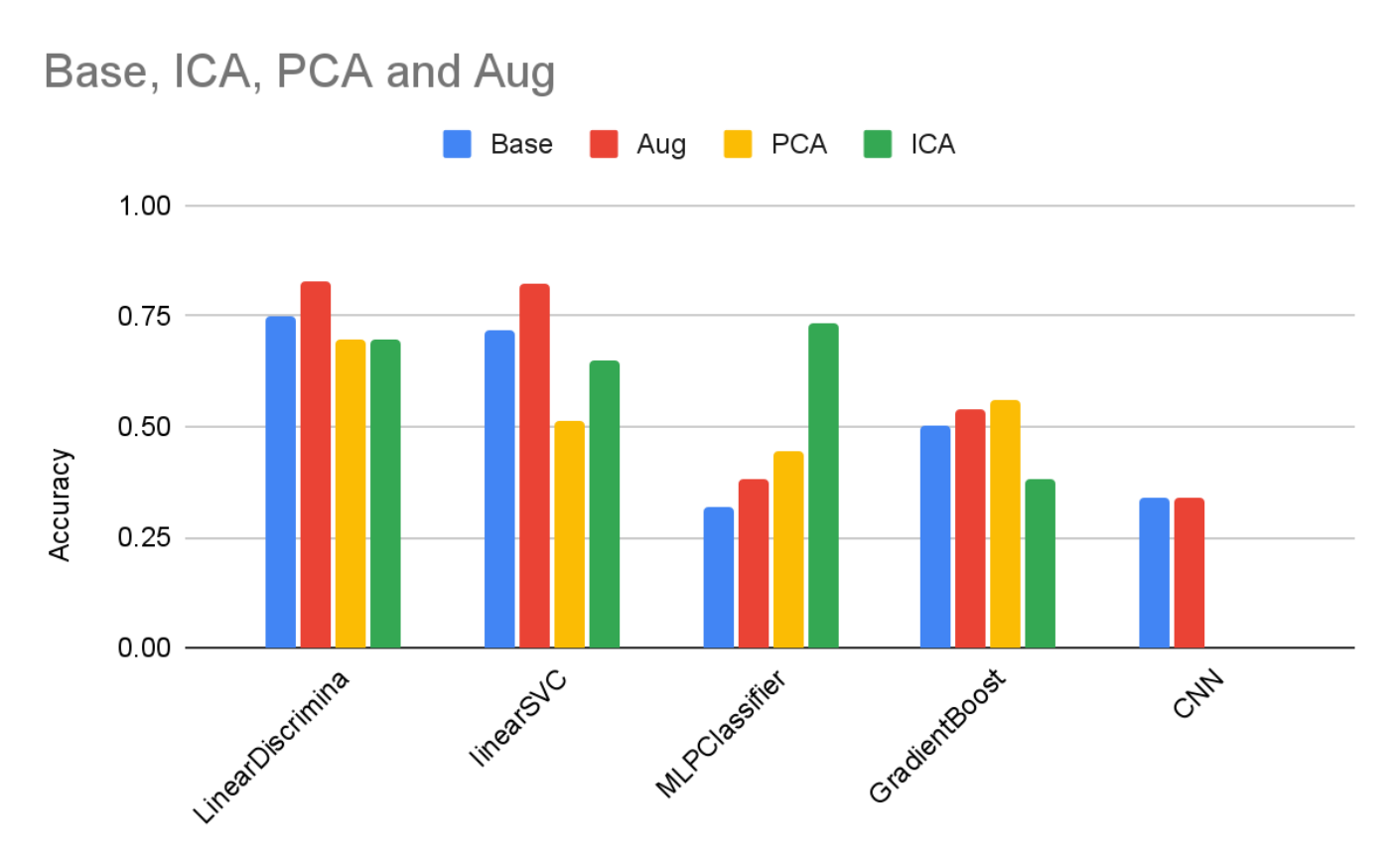}
    \caption{Results of classifiers on with different pre-processing methods.}
    \label{fig:figRobert2}
\end{figure}

\begin{figure}[ht]
    \centering
    \includegraphics[scale = 0.75]{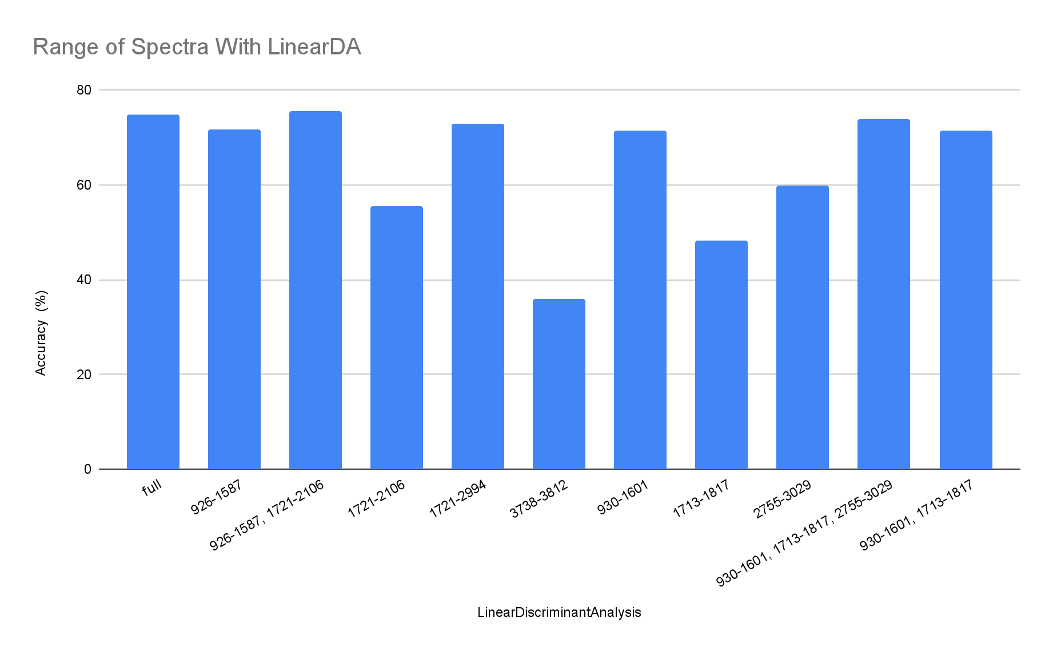}
    \caption{Results of Linear Discriminant Analysis for different feature selection.}
    \label{fig:figRobert3}
\end{figure}

In the second stage, the classifiers were tested in conjunction with PCA, ICA (\texttt{scikit-learn} methods of PCA and FastICA \citep{pedregosa2011scikit} were used) or data augmentation. The use of PCA and ICA altered the data by reducing the dimensionality, while on the other hand data augmentation increases the number of samples. For data augmentation, the data augment function from \citet{bjerrum2017data} was used. This increased the number of training samples from 3,244 to 19,464. At this stage, only a subset of the previously tested model were considered, based on their performances in the previous step. Figure \ref{fig:figRobert2} shows the results of each classifier with each pre-processing method (base, ICA, PCA, data augmentation (Aug)). From these results, it was noted that LDA following data augmentation achieved the highest accuracy with 82.7\%. The greatest improvement in the predictions was observed using MLP after ICA (improvement of 41\%). An additional experiment was then carried out with just the use of the LDA model. This was to show the importance of regions within the spectra, and a number of different wavelength region were tested. Therefore, figure \ref{fig:figRobert3} shows the results of the LDA when removing different spectral regions.

There was a general increase in accuracy over the base approach when data augmentation was used, with the only exception of CNN. With regard to wavelengths selection, there was no noticeable increase in accuracy when focusing on a specific region in the spectra. Nonetheless, the majority of the relevant information lied within the region from 925 cm$^{-1}$ and 1597 cm$^{-1}$, and there was a slight increase in the accuracy of prediction of around 1\% when using the range of 925 to 1585 cm$^{-1}$ and 1717 to 2103 cm$^{-1}$ compared to the full set of wavelengths. 

\subsection{Participant 5}\label{sec:suzyJoe}

In order to prepare the data set for predictive analysis, some pre-processing was considered. As directed by the challenge organisers, outlying spectra were removed such that the data set consisted of 3243 transmittance spectra covering 1060 wavelengths. Subsequently, spectra were transformed to absorbance values by taking $\log_{10}$ of the reciprocal of the transmittance values. In addition, following \cite{frizzarin2021application}, a subset of 534 wavelengths that lay outside the water-related high-noise-level regions were identified as relevant for predicting a cow's diet, although the water-regions were not excluded at this point in the analysis. 

To ensure a robust assessment, the dataset was split into training and validation sets. In this case, the validation set was constructed to control for batch effect confounding, which may bias estimates for out-of-sample prediction \citep{soneson2014batch}. Inspection of the data set revealed that rows were ordered to have several consecutive observations of each diet. Therefore, it was assumed that each set of consecutive diet observations belonged to a single batch. In this manner, 90 batches, 30 for each diet, were identified. In addition, the data was collected over three years \citep{frizzarin2021application}, and so it was assumed that the first 30 batches were collected in the first year of the study, the next 30 in the second year, and the final 30 in the third. Based on these assumptions, the validation set consisted of 996 spectra from 30 batches collected in the study's third year, which included ten batches for each diet, while models have been trained on the 2247 remaining spectra. Training data was randomly split into $ V =10$ folds, with each fold including two batches from each diet. Possible batch effect of repeated measurements for a single cow were ignored.

In order to describe the predictive model used in this analysis, let $\mathcal D = \left\{ y_i, \mathbf x_i \right\}_{i = 1}^N$ denote the observed data, where the response variable $y_i \in \left\{ 1, \dots, M \right\}$ represents the diet of the $i$-th cow and covariates $\mathbf x_i \in \mathbb R^D$ represent the corresponding milk absorbance spectrum. Note that this analysis considers $M = 3$ diets, $D = 1060$ wavelengths, and $N = 3243$ training observations. The objective of the proposed predictive models is to learn $\mathbb{P} \left( y \mid \mathbf x \right)$, that is, the probability that a given milk sample comes from a grass, clover or TMR-fed cow, given the spectrum for that sample. 

The first step in constructing a predictive model is to define a deterministic mapping function $g : \mathbf x_i \to \mathbf z_i$, for $\mathbf z_i \in \mathbb R^{D'}$, with $D' < D$, which describes a feature extraction procedure. Two approaches to feature extraction were considered here. The first simply selected the $D' = 534$ relevant wavelengths identified by \cite{frizzarin2021application} such that $\mathbf z_i$ is the $i$-th absorbance spectrum after removing the high-noise-level water regions and standardises each wavelength. The second was based on the wavelet transform, a popular technique for signal processing which can be applied for data compression, smoothing, and multi-resolution analysis \citep{nason2008wavelet}, and proceeds in three steps. After setting high-noise-level regions of each spectrum to 0, a thresholded wavelet transform provides a set of wavelet coefficients. The feature vector $\mathbf z_i$ is then the vector of wavelet coefficients that are non-zero for at least one of the $N$ spectra, in this case $D' = 594$.
The thresholded wavelet transform is available with the \texttt{wavethresh} R package \citep{nason2016wavethresh}, using Daubechies least symmetric wavelet as the mother wavelet and Bayesian approach to thresholding wavelet coefficients \citep{abramovich1998wavelet}.
Note that setting wavelengths in the high-noise-level regions to 0 means the wavelet transform preserves the spectral distance between wavelengths while ensuring that the corresponding wavelet coefficients are 0. 

Given the feature vector $\mathbf z_i = g \left( \mathbf x_i \right)$, a multinomial regression model for diet was assumed, such that
\begin{equation}
    \mathbb{P} \left( y_i = m \mid \mathbf z_i \right) = \frac{\exp \left(\boldsymbol \beta_m^\top \mathbf z_i \right)}{\sum_{l = 1}^{M} \exp \left(\boldsymbol \beta_l^\top \mathbf z_i \right)},
\end{equation}
for $m = 1, \dots M$ where $\boldsymbol \beta_m \in \mathbb R^{D'}$, implicitly assuming that $\mathbf z_i$ includes an intercept term. The \texttt{glmnet} package \citep{friedman2010regularization} fits this model to data efficiently. For simplicity, a LASSO model was fitted, where 10-fold cross-validation on the training data informs the penalty hyperparameter.

Finally, the predictive performance of the proposed models was compared by analysing their log-loss on the validation data set. That is, for a validation data set and a model $\mathcal M_j$ for $\mathbf z_i = g \left( \mathbf x_i \right)$, the log-loss is defined as
\begin{equation}
    \ell_j = - \frac{1}{N'} \sum_{i = 1}^{N'} \sum_{m = 1}^M \mathbb{I} \left( y_i = m \right) \ln \mathbb{P} \left( y_i = m \mid \mathbf z_i, \mathcal M_j \right),
\end{equation}
where $N'$ is the number of observations in the validation set, $\mathbb{I} \left( \mathcal{A} \right)$ is the usual indicator function that is equal to 1 when $\mathcal{A}$ is true and 0 otherwise and $\mathbb{P} \left( y_i = m \mid \mathbf z_i, \mathcal M_j \right)$ is the probability under $\mathcal M_j$ that $y_i = m$ given $\mathbf z_i$. The log-loss is a proper scoring rule for evaluating predictive models \citep{gneiting2007strictly}, where smaller scores are better, and so encourages to express the true belief about the data. It is also straightforward to set benchmarks for assessing the quality of predictions a priori. For example, for any $M$ a mean log loss of 0 represents perfect predictive performance, while when $M = 3$ as in the considered case, a mean log loss of $-\ln (1 / 3) \approx 1.1$ represents ``guessing'', where we predict each category uniformly at random. For completeness, the classification accuracy of $\mathcal M_j$ was also assessed.

The results of this analysis are presented in Table \ref{tab:results}. The first model considered was a LASSO-penalized multinomial regression of the raw milk spectra on the diet, where high-noise-level regions of the spectrum was excluded and the wavelengths standardised. The tuning parameter $\lambda$, controlling the strength of the penalization, was selected to minimise the multinomial deviance (a statistic proportional to the mean log-loss) via 10-fold cross-validation. 
The log-loss of this model on the training set was 0.57, which corresponds to a diet classification accuracy of 77\%. A closer examination of the predictions revealed that when \texttt{CLV} and \texttt{GRS} were treated as a single category (pasture-fed), it was possible to predict \texttt{TMR} with an accuracy of 94\%. When trying to predict whether the cow was fed \texttt{CLV}, given that it was pasture-fed, an accuracy of 72\% was achieved. Predictive performance was much poorer on the validation set, with an overall log-loss of 0.82, corresponding to an accuracy of 58\%. The model predicted \texttt{TMR} with an accuracy of 88\%. However, for cows known to be pasture-fed, it predicted \texttt{CLV} with an accuracy of 49\%. 

The second model considered a multinomial regression of the non-zero thresholded wavelet transform coefficients of the milk spectra on diet. As above, the model was fitted by maximising a penalised log-likelihood and by using 10-fold cross-validation to tune $\lambda$. For this model, the log-loss on the training set was equal to 0.74, corresponding to an accuracy of 69\%, although it predicted \texttt{TMR} with an accuracy of 88\%. For pasture-fed cows, it predicted \texttt{CLV} with an accuracy of 68\%.
As with the first model, performance dropped for the validation set. The log-loss was 0.88 and \texttt{TMR} accuracy was 79\%. Given that a cow was pasture-fed, the \texttt{CLV} accuracy was 47\%. These results are summarised in Table \ref{tab:results}.

The obtained results clearly showed that milk spectra carry a signal distinguishing pasture-fed cows from \texttt{TMR}, but that it was difficult to distinguish between \texttt{CLV} and \texttt{GRS}. However, the predictive performance was much poorer on the validation dataset than for the training one, indicating that the adopted models did not offer a robust out-of-sample predictions. Without careful consideration of potential batch effect confounders within the sampled spectra, we are likely to overestimate the out-of-sample performance of our models. Collecting data from more cows over a more extended period should alleviate this issue and allow more robust models to be developed.

Lastly, no evidence was found to suggest that wavelet transformed spectra provided helpful insight into the cows' diet. However, that is not to say that some alternative basis expansion could improve the current predictive models. In fact, given more data on the relationship between milk spectra and diet, the development of models which allow for non-linear relationships between wavelengths may prove a fruitful avenue for future research.

\begin{center}
\begin{table}[]
    \caption{Predictive model assessment.}
    \centering
    \begin{tabular}{ c|c c } 
    Model & In-sample log-loss & Validation log-loss \\
    \hline
    Raw Spectra & 0.57 & 0.82 \\ 
    \hline
    Wavelet Coefficents & 0.74 & 0.88 \\
    \end{tabular}
    \label{tab:results}
\end{table}
\end{center}

\subsection{Participant 6}\label{sec:katarina}

As a first step, the training set was centered and scaled and the same transformation was applied to the test set. In the following analyses, no outliers were removed while all the spectra were transformed from transmittance to absorbance. Wavelengths from high-noise level spectral regions between 1720 and 1592 cm$^{-1}$, between 3698 and 2996 cm$^{-1}$, and greater than 3,818 cm$^{-1}$ were removed from the analysis following \citet{frizzarin2021application}. 

\begin{figure}[t]
    \centering
    \includegraphics[width = 15cm]{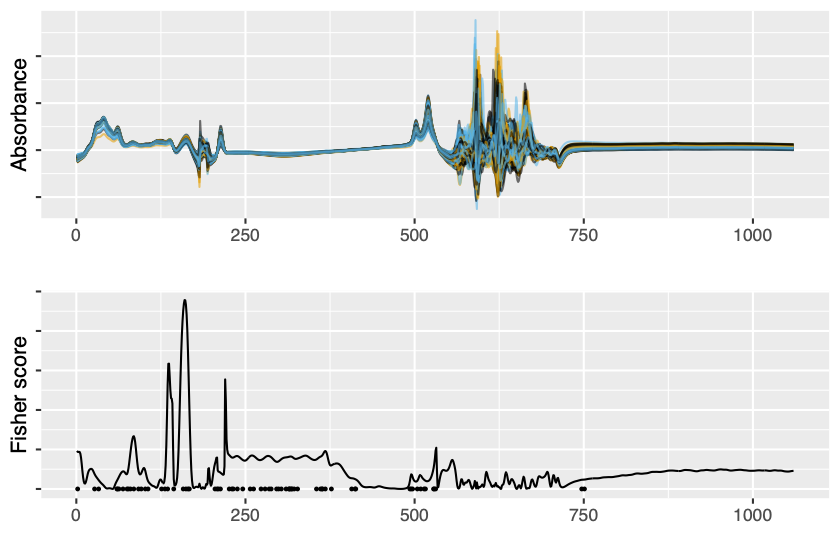}
    \caption{Spectra absorbance and the corresponding Fisher score with points on the x-axis denoting the wavelengths selected by the GA.}
    \label{fig:katarina}
\end{figure}

The Fisher score, being the ratio of between to within diet group variance, was calculated for all the wavelengths in the training set. For wavelength $j$, the Fisher score is given by:
$$
\text{Fisher score}_j = \frac{\sum_{m = 1}^M\sum_{i = 1}^n \mathbb{I}(y_{i} = m)(\bar{x}_{\cdot j}^{(m)} - \bar{x}_{\cdot j})^2}{\sum_{m = 1}^M\sum_{i = 1}^n \mathbb{I}(y_{i} = m)(x_{ik}- \bar{x}_{\cdot j}^{(m)})^2}
$$
where $j$ denotes the wavelength index, $i = 1,\dots,n$ denotes the spectra with $n$ being the number of spectra in the training set, $m$ denotes the diet group with $M = 3$, $\mathbb{I}(y_{i} = m)$ is an indicator of diet group spectra $i$, $\bar{x}_{\cdot j}$ is the average of wavelength $j$ for all spectra $(i = 1, \dots, n)$, $\bar{x}_{\cdot j}^{(m)}$ is the average of wavelength $j$ in diet group $m$. A wavelength with the highest Fisher score in each of the discarded regions was kept in the analysis. Wavelengths with Fisher score lower than 0.002 were removed from further analysis, thus leaving 380 wavelengths. In order to compare algorithms and carry out further feature selection, the training set was itself randomly split 75/25 into training and testing sets stratified by diet. A genetic algorithm \citep{holland1992adaptation}, implemented in library \texttt{genalg} \citep{willighagen2022} was used as a stochastic search method to find an optimal subset of input wavelengths for classification. Individuals in the GA population were represented by binary strings denoting wavelengths to be included or excluded for prediction. Objective function was set to be the average accuracy from ten cross-validated fits of linear discriminant analysis (LDA) of the training subset. GA was run for 200 iterations with population size set at 200. Figure \ref{fig:katarina} shows the spectra absorbance and the corresponding Fisher scores, with points denoting the wavelengths selected by the GA. 

The best configuration from the final GA population had 70 wavelengths included. These wavelengths were used as inputs to the following classification algorithms:
\begin{itemize}
    \item Linear discriminant analysis (LDA), library \texttt{MASS} \citep{MASSpackage}; 
    \item Partial least squares discriminant analysis (PLS-DA) \citep{plspackage};
    \item Least absolute shrinkage and selection operator \citep[LASSO;][]{tibshirani1996regression}, library \texttt{glmnet} \citep{friedman2010regularization}; 
    \item Elastic net \citep[EN;][]{zou2005regularization}, library \texttt{glmnet};
    \item Random Forest \citep[RF;][]{breiman2001random}, library \texttt{ranger} \citep{ranger}; 
    \item Support vector machines \citep{vapnik1998}, library \texttt{kernlab} \citep{kernlab}; 
    \item Bayesian kernel projection classifier \citep[BKPC][]{domijan2011bayesian}, library \texttt{BKPC} \citep{bkpc}.
\end{itemize}

All analyses were done using \texttt{R} \citep{R}, the code is available in the github repository \url{https://github.com/domijan/KD\_Vistamilk2022}. 

The training set was randomly split into ten further training/testing sets of equal size, stratified on diet. The average accuracy and standard deviation over the ten random splits for all the classification algorithms are given in Table \ref{tab:tableKatarina}. LDA performed best with average accuracy of 77.4\%. PLS-DA and EN overall accuracy was of 76.9\%, 76.5\% respectively. The algorithms were tuned using further cross-validation of the training sets.  For BKPC and SVM, the best results were obtained with a linear kernel. The predictions of the LDA were submitted to the competition. Moreover, genetic algorithm was able to select a much smaller subset of wavelengths without loss of classification performance. 

\begin{table}[t]
\centering
\caption{Average accuracy for over ten random splits of the training set for classifiers. LDA: linear discriminant analysis; PLS: partial least squares regression; EN: elastic net; BKPC: Bayesian kernel projection classifier; SVM: support vector machine; LASSO: Least absolute shrinkage and selection operator; RF: random forest.}
\begin{tabular}{l|ccccccc}
Accuracy & LDA & PLS & EN & BKPC & SVM & LASSO & RF \\
\hline
Mean     & 0.774        & 0.769        & 0.765       & 0.759         & 0.738        & 0.736          & 0.509       \\
SD      & 0.008        & 0.009        & 0.007       & 0.008         & 0.007        & 0.006          & 0.014      
\end{tabular}
\label{tab:tableKatarina}
\end{table}

\section{Discussion}\label{sec:sec4}

While the dataset provided for the data competition included three different classes to discriminate (i.e. TRM, GRS, and CLV), the main difficulty of the present data competition was concerned with the discrimination between GRS and CLV diets. In fact, the ability of distinguishing pasture and TMR dietary regimens has been already documented \citep{frizzarin2021application}, with the discrimination being driven mainly by the different content of fatty acids (FA) in milk \citep{agradi2020determination}. In particular, milk from pasture based diet is generally richer in saturated FA such as linoleic acid, poorer in saturate FA, and have a lower omega6/omega3 ratio \citep[see e.g.][]{chilliard2007diet,dewhurst2006increasing,ferlay2013,ferlay2017production}. As MIR is known to be able to predict, with a certain degree of accuracy, the different FA in milk \citep{soyeurt2011mid}, spectral data are therefore capable to discriminate also TMR and pasture diets.

On the other hand, since GRS and CLV dietary regimens differed only for the inclusion of 20\% annual clover in perennial ryegrass sward for the CLV diet, induced differences in the FA might be less clear. As a consequence, to discriminate GRS and CLV exploiting spectral information only, a careful and accurate tuning of the modelling choices was required. In this regard, interestingly, some participants proposed two-steps classification approaches, with the first step focusing on TMR and pasture based diets, while the second one aimed at distinguishing CLV from GRS samples. As an example, participant 2 highlighted a potentially significant gain in terms of accuracy when considering an ensemble approach, where components extracted from LDA was used to train a linear SVM, better discriminating between GRS and CLV. Again, in Section \ref{sec:memb1_unibo} two consecutive LDA models have been fitted, with the first one being used to discriminate TMR from pasture while the second, exploiting the discriminant function on the pasture samples only, was trained to classify GRS and CLV. 

Generally speaking, linear approaches introduce a gain in interpretability of the results, while paying a price in terms of accuracy. Nonetheless, the review of the different approaches presented in this paper showed that strong performances were achieved resorting to linear classifiers. In fact, remarkable results were obtained when adopting LDA-based approaches (see, e.g., participants 1, 2, 4 and 6), which were certainly proven effective in discriminating TMR and pasture diets and, as highlighted above, were also used as a building block for promising two-steps procedures. Nevertheless, the approaches presented in Sections \ref{sec:tach} and \ref{sec:deepLearningGeorg}, which attained the best test set prediction accuracies as it is displayed in Table \ref{tab:finalAccuracies}, pointed towards the need of considering non-linearities, especially when the aim is to discriminate between GRS and CLV. This is confirmed by the confusion matrix displayed in Table \ref{tab:tach_confusion}, where it is shown that these two different dietary regimens are discriminated remarkably well, especially if considering their similarities from a compositional standpoint. Note that, while with FCN interpretation of the results and exploration of the most informative wavelengths are compromised, the approach in Section \ref{sec:tach}, which is considering again LDA as the final classifier, tends to be more transparent. However, the clever random polynomial variables generation proposed tends to produce new features which are difficult to interpret from a chemical standpoint. Therefore, as it often happens in modern data analysis routine, the adopted approaches have to be tailored on the specific aim to pursue, often dealing with the standard trade-off between accuracy and interpretability. 
\begin{center}
\begin{table}[t]
    \caption{Accuracy computed on the test dataset for all the participants.}
    \centering
    \begin{tabular}{c|cccccc} 
    Participant & Sect \ref{sec:tach} & Sect \ref{sec:deepLearningGeorg} & Sect \ref{sec:hussein} & Sect \ref{sec:memb1_unibo} & Sect \ref{sec:memb2_unibo}   \\
    \hline
    Test accuracy & 0.871 & 0.837 & 0.798 & 0.711 &  0.783   \\ 
    \hline
    \hline
    Participant & Sect \ref{sec:memb3_unibo} & Sect \ref{sec:robertwilliamson} & Sect \ref{sec:suzyJoe}  & Sect \ref{sec:katarina} & &  \\
    \hline
    Test accuracy & 0.796 & 0.786 & 0.724 & 0.766 &  \\ 
    \end{tabular}
    \label{tab:finalAccuracies}
\end{table}
\end{center}

Data transformation is widely used in near-infrared analyses, as the analysed samples are generally more noisy. Differently, samples analysed using MIRS are generally less noisy, therefore these transformations, with the exception for the transformation of the wavelengths from transmittance to absorbance, are not widely used. In the present study some data transformations were tested, but the reported results confirmed that they do not have a strong impact on the quality of the prediction results. Differently from data transformation, the removal of the spectral regions related to water is of fundamental importance, as reported by the participants which tested their prediction methods before and after their removal. For example, results from Section \ref{sec:tabular} showed an improvement of 11.6\% and of 25.7\% when ridge regression and LDA were respectively used in combination of new polynomial variables generation after water regions removal. Again, in Section \ref{sec:deepLearningGeorg} an improvement of the prediction performance, from 17.5\% (CNN) to 20.5\% (FCN), after removing the water regions also when using deep learning methods is shown. Participant 1 also demonstrated the possibility to select the important variables directly from the spectra, in fact they achieved the best prediction results using a variables selection approach starting from all the spectral information (see Table \ref{tab:tabular_results}). Variable selection was also tested in Section \ref{sec:katarina}, where a genetic algorithm was used to select a smaller subset of wavelengths without substantial loss in classification performance.

\begin{table}[t]
\caption{Final confusion matrix obtained with the approach outlined in Section Sect \ref{sec:tach}.}
\centering
\begin{tabular}{rr|ccc}

  & & & Actual &  \\ 
 & & CLV & GRS & TMR \\ 
  \hline
& CLV & 312 &  55 &   5 \\ 
Predicted & GRS &  61 & 300 &   5 \\ 
 & TMR &   6 &   7 & 326 \\ 
   \hline
\end{tabular}
\label{tab:tach_confusion}
\end{table}

In Section \ref{sec:group_bologna}, the participants investigated the pairwise agreement among the three different approaches, to calculate by comparing the observations and quantifying the percentage of classifications in agreement on the total number of observations (Table \ref{tab:unibo}). Methods applied by members 1 and 2 gave similar predictions (agreement of 84.21\%), whereby agreement between predictions from member 3 was between 70.84\% (with member 2) and 72.90\% (with member 1). Although strong, the discrepancies among the three predictions could be due to: i) the different number of samples retained for model development, and ii) the different number of predictors (i.e., wavelengths) used for training, considering that the first member used the entire edited spectra, whereby the second and third applied different algorithms for wavelengths selection. This investigation from the third participant permits to understand that differences in data editing and different methodologies selected for the predictions, even if similar, brought to consistently different class predictions.

A final discussion point was related to the creation of the test dataset. The dataset was created by the organizers, who splitted the original dataset in 75\% training and 25\% test dataset, considering a correct division of the classes across years into the 2 datasets. The discussion revolved around whether or not divide the dataset into 75\% training and 25\% testing, or dividing the dataset according to time components, like keeping the samples recorded in 2015 and 2016 into the training dataset, and the samples recorded in 2017 in the test dataset. Such temporal division would permit to understand if samples recorded in previous years can predict future information.

\section{Conclusion}\label{sec:conclusion}

Thanks to the high number of participants, with different backgrounds, who provided their prediction results, the data competition was a thought-provoking occasion to discuss some of the challenges arising when analyzing spectral data and provided insightful indications. 

As mentioned in the paper and as it was previously shown in \citet{frizzarin2021application}, the stronger compositional dissimilarities between pasture-based diet and TMR-based ones induced an easier discrimination between the corresponding classes. This generally led to overall good performances, in terms of accuracy, for the adopted methods (see Table \ref{tab:finalAccuracies}). On the other hand, the distinction between milk samples originated from GRS and CLV was more challenging. Nonetheless, as it is shown in Table \ref{tab:tach_confusion}, some hand-crafted strategies specifically proposed for this competition showed more than promising results also when employed to detect differences in the composition between distinct pasture-based feeding regimens. 
In particular, non-linear transformations of the original wavelengths and two-steps classification approaches, outlined in Section \ref{sec:tabular} and \ref{sec:group_bologna}, seemed to be effective in solving this problem. 

Pre-treatments were generally not beneficial for the improvement of the prediction equations, while the deletion of the spectral regions related to water (with manual selection of these regions or by means of automatic variable selection procedures) improved the prediction results. The utilization of linear models, in particular LDA, provided some of the best results, and the overall best prediction was achieved using LDA applied after wavelengths selection and random polynomial generation, as it was shown in Table \ref{tab:finalAccuracies}. When spectral analyses are undertaken it is important to know not only the best possible statistical methods to use for the analyses, but also what is the best data editing for such data.

\newpage

\appendix
\section{Supplementary material}\label{sec:supplMat}

\subsection{Deep neural network architecture}\label{sec:appendixA}

\begin{table}[h]
\centering
\caption{List of the deep model architectures considered in Section \ref{sec:deepLearningGeorg}, including the number of trainable parameters for each model and the type of input data they accept.}
\begin{tabular}{lll}
\hline
\textbf{Model Architecture}    & \textbf{Parameters} & \textbf{Input Data and Shape}                                                                                   \\ \hline
\begin{tabular}[c]{@{}l@{}}\textbf{FCN}\\ - Dense layers of 1024, 512, 128, 64 and 32 units\\ - Output layer of 3 units\\ - Dropout for dense layers, drop rate of 0.2\\ - elu activation for hidden layers\\ - softmax activation for output layer\\ - Adam optimiser, initial learning rate of 0.0001\\ - Categorical cross entropy as loss function\end{tabular}                                                                           & 1,785,923  & \begin{tabular}[c]{@{}l@{}}- Linear, full (1060)\\ - Linear, reduced (518)\end{tabular}      \\ \hline
\begin{tabular}[c]{@{}l@{}}\textbf{CNN}\\ - Convolutional layers with 32, 64 and 128 filters\\ - Filters of shape (3, 3), (2, 2) and (2, 2)\\ - Flattening layer\\ - Dense layers of 512, 256, 128, 64, and 32 units\\ - Output layer of 3 units\\ - elu activation for hidden layers\\ - softmax activation for output layer\\ - Adam optimiser, initial learning rate of 0.0001\\ - Categorical cross entropy as loss function\end{tabular} & 55,332,419 & \begin{tabular}[c]{@{}l@{}}- Squared, full (33x33)\\ - Squared, reduced (23x23)\end{tabular} \\ \hline
\begin{tabular}[c]{@{}l@{}}\textbf{CNN\_DILATED}\\ - Same architecture as CNN\\ - Kernels built with a dilation rate of (2, 2)\end{tabular}                                                                                                                                                                                                                                                                                                   & 41,176,643 & \begin{tabular}[c]{@{}l@{}}- Squared, full (33x33)\\ - Squared, reduced (23x23)\end{tabular} \\ \hline
\end{tabular}
\label{deeparch}
\end{table}

\newpage
\subsection{Participant 3}\label{sec:gruppoUniBo}

\begin{table}[ht]
    \centering
    \caption{Standardized canonical discriminant function coefficients of the variables selected by DA and effective size measures.}
\begin{tabular}{lll}
\hline
\textbf{Wavenumber, cm$^{-1}$} & \textbf{Function} & \\
 & 1 & 2 \\
\hline 
1069 & 2.899 & 0.298 \\
1130 &	-3.790 &	0.416 \\
1181 &	-2.003 &	5.371 \\
1269&	-7.321	&-2.495 \\
1292&	10.544	&-3.045 \\
1377	&-5.860& -0.482 \\
1416&	-5.885&	1.267 \\
1439&	12.710&	1.112 \\
1474&	-4.689&	3.714 \\
1539&	-3.816&	-2.385 \\
1577&	4.442&	1.247 \\
1752&	11.958&	6.035 \\
2782&	-1.459&	0.875 \\
2851&	-15.686&	-13.612 \\
2890&	16.085&	3.459 \\
2932&	-4.166&	0.916 \\
\hline
\textbf{Eigenvalue} & 1.732 & 0.109 \\
\hline
\textbf{\& of variance} & 94.1\% & 5.9\% \\
\hline
\textbf{Canonical correlation} & 0.796 & 0.313 \\
\hline
\end{tabular}
    \label{tab:uniBoAppendix}
\end{table}

\begin{table}[h]
    \centering
\caption{Group means (centroids) for the Discriminant Functions}
\begin{tabular}{l|ll}
\hline
\textbf{Diet} & \textbf{Function} & \\
 & 1 & 2 \\
\hline 
CLV & 0.872 & 0.403 \\
GRS	& 0.954 & -0.400 \\
TMR & -1.895 & -0.012 \\
\hline
\end{tabular}
    \label{tab:uniBoAppendix2}
\end{table}

\begin{table}[]
\caption{Classification related statistics and leave-one-out cross-validation. $^a$ 71\% of original grouped cases correctly classified. $^b$ Cross-validation is done only for those cases in the analysis. In cross-validation, each case is classified by the functions derived from all cases other than that case. 70.5\% of cross-validatd grouped cases correctly classified.}
\begin{tabular}{lll|cccc}
\hline
\textbf{}  & \textbf{} & \multirow{2}{*}{\textbf{Diet}} & \multicolumn{3}{c}{\textbf{Predicted Group Membership}} & \multirow{2}{*}{\textbf{Total}} \\
\textbf{}  & \textbf{} &  & \textbf{CLV} & \textbf{GRS} & \textbf{TMR}  &    \\
\hline
\multirow{6}{*}{\textbf{Original$^a$}}  & \multirow{3}{*}{\textbf{Count}} & \textbf{CLV}  & 629 & 363  & 83 & 1075  \\                                       &   & \textbf{GRS} & 323 & 668   & 62    & 1053  \\
&  & \textbf{TMR}  & 39  & 44  & 942  & 1025  \\
\cline{2-7}
& \multirow{3}{*}{\textbf{\%}}  & \textbf{CLV}  & 58.5  & 33.8  & 7.7 & 100.0 \\
 & & \textbf{GRS} & 30.7 & 63.4 & 5.9 & 100.0  \\
 &  & \textbf{TMR}  & 3.8  & 4.3  & 91.9 & 100.0   \\
\hline
\multirow{6}{*}{\textbf{Cross-validated$^b$}} & \multirow{3}{*}{\textbf{Count}} & \textbf{CLV}                   & 620               & 369              & 86               & 1075                            \\
                                           &                                 & \textbf{GRS}                   & 326               & 663              & 64               & 1053                            \\
                                           &                                 & \textbf{TMR}                   & 39                & 47               & 939              & 1025                            \\
                                           \cline{2-7}
                                           & \multirow{3}{*}{\textbf{\%}}    & \textbf{CLV}                   & 57.7              & 34.3             & 8.0              & 100.0                           \\
                                           &                                 & \textbf{GRS}                   & 31.0              & 63.0             & 6.1              & 100.0                           \\
                                           &                                 & \textbf{TMR}                   & 3.8               & 4.6              & 91.6             & 100.0                          
\end{tabular}
\label{tab:uniBoAppendix3}
\end{table}

\begin{figure}[h]
  \centering
  \includegraphics[scale = 0.5]{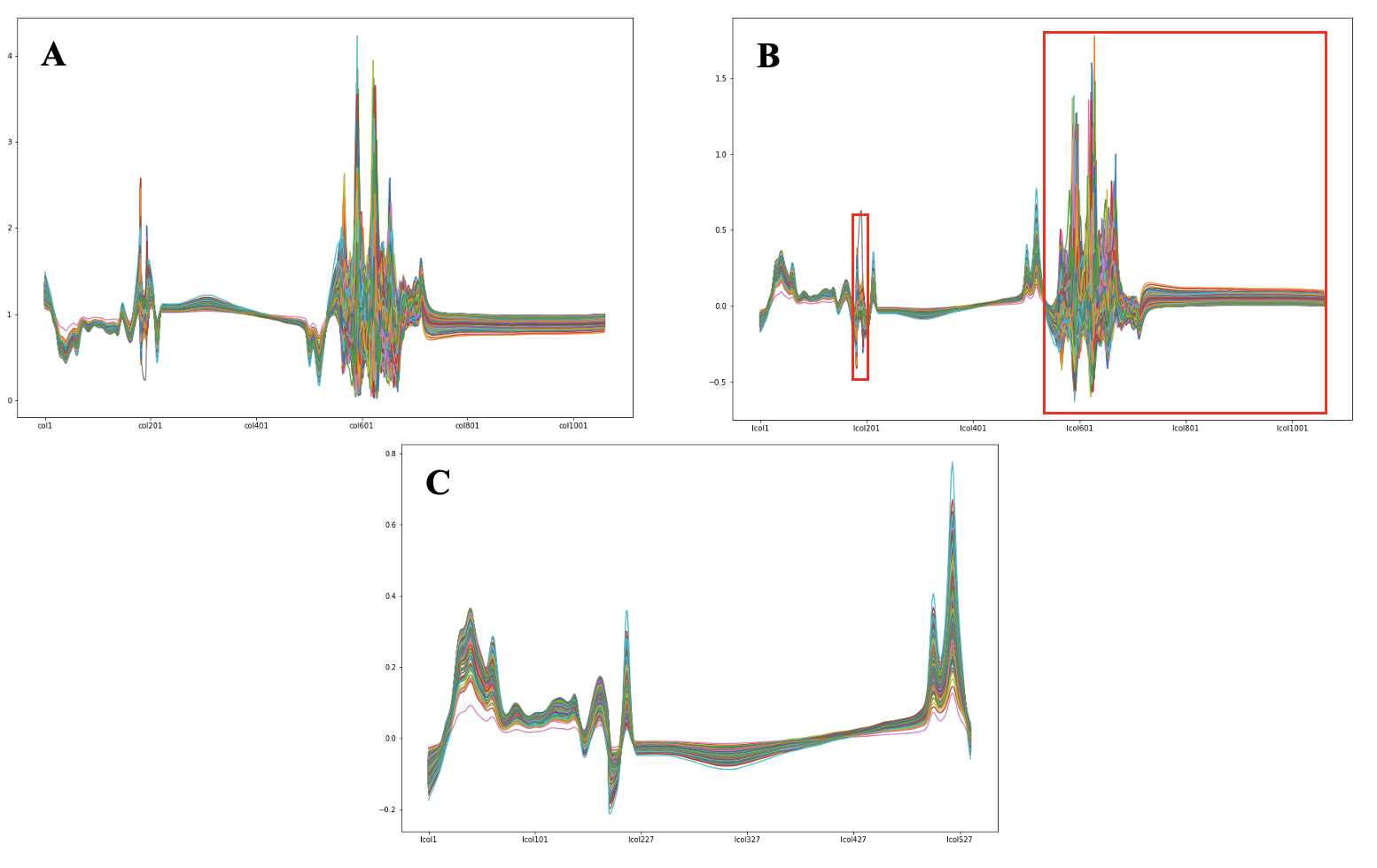}
  \caption{Line plot of raw spectra expressed in transmittance (A), conversion of raw spectra from transmittance fo absorbance (B; red boxes indicate low signal-to-noise regions), and raw spectra in absorbance after noisy area removal (C). }
  \label{fig:figUniBoSupplementary} 
\end{figure}

\newpage 
\section*{Acknowledgements}
This publication has emanated from research conducted with the financial support of Science Foundation Ireland (SFI) and the Department of Agriculture, Food and Marine on behalf of the Government of Ireland under grant number (16/RC/3835), the SFI Insight Research Centre under grant number (SFI/12/RC/2289\_P2) and the SFI Starting Investigator Research Grant ``Infrared spectroscopy analysis of milk as a low-cost solution to identify efficient and profitable dairy cows'' (18/SIRG/5562).

\section*{Declaration of interests}
The authors declare that they have no known competing financial interests or personal relationships that could have appeared to influence the work reported in this paper. 
\bibliographystyle{apalike}
\bibliography{biblio.bib}

\end{document}